\documentclass[11pt,a4paper]{article}

\usepackage[a4paper,margin=2.15cm]{geometry}
\usepackage{setspace}
\usepackage{amsmath,amssymb,amsthm,mathtools}
\usepackage{bm}

\usepackage{booktabs}
\usepackage{array}
\usepackage[dvipsnames]{xcolor}
\usepackage{graphicx}
\usepackage{caption}
\usepackage{subcaption}
\usepackage{tabularx}
\usepackage{booktabs}

\usepackage{comment}

\usepackage[colorlinks=true,linkcolor=blue,citecolor=blue,urlcolor=blue]{hyperref}

\usepackage[numbers,sort&compress]{natbib}

\numberwithin{equation}{section}

\renewcommand{\d}{{\rm d}}

\newtheorem{theorem}{Theorem}

\newtheorem{prop}{Proposition}
\newtheorem{lemma}{Lemma}
\newtheorem*{conjecture}{Conjecture}

\theoremstyle{definition}

\theoremstyle{remark}
\newtheorem{remark}{Remark}

\newcommand{\R}{\mathbb{R}}
\newcommand{\C}{\mathbb{C}}
\newcommand{\N}{\mathbb{N}}
\newcommand{\eu}{\ensuremath{\mathrm{e}}}

\def\D{{\rm D}}

\usepackage{tikz}
\usetikzlibrary{decorations.pathmorphing}

\newcommand{\spring}{%
  \mathord{\tikz[baseline=-0.5ex,x=1em,y=1em]{
    \draw (0,0)--(0.2,0);
    \draw[decorate,decoration={zigzag,
      segment length=3pt,amplitude=2pt}]
      (0.2,0)--(1.2,0);
    \draw (1.2,0)--(1.4,0);
  }}%
}

\newcommand{\springpot}{%
  \mathord{\tikz[baseline=-0.5ex,x=1em,y=1em]{
    \draw (0,0)--(0.3,0);
    \draw (0.3,0)--(0.7,0.3)--(1.1,0)
      --(0.7,-0.3)--cycle;
    \draw (1.1,0)--(1.4,0);
  }}%
}

\title{\bf A Representation Theorem for\\ 
Fractional Viscoelastic Networks}

\author{
A.~Giusti$^{abc}$\thanks{Email: andrea.giusti9@unibo.it},
$\ $
I.~Colombaro$^{bd}$,
$\ $
A.~Mentrelli$^{bce}$,
$\ $
F.~Mainardi$^{a}$
$\ $
\\
\\
$^a${\em DIFA, University of Bologna, 40126 Bologna, Italy}
\\
\\
$^b${\em Alma Mater Research Center on Applied Mathematics $({\cal AM}^2)$,}\\ 
{\em University of Bologna, I-40123 Bologna, Italy}
\\
\\
$^c${\em I.N.F.N., Sezione di Bologna, I.S.~FLAG, 40127 Bologna, Italy}
\\
\\
$^d${\em Faculty of Engineering, Free University of Bozen-Bolzano, 39100 Bolzano, Italy}
\\
\\
$^e${\em Department of Mathematics, University of Bologna, 40123 Bologna, Italy}
}

\date{ }

\begin{document}

\maketitle

\begin{abstract}
\noindent We prove a representation theorem for finite series–parallel viscoelastic networks composed of springs and spring-pots of a common order $0<\alpha<1$. Using classical representation results for complete Bernstein and Stieltjes functions, we show that the associated complex shear modulus, expressed in an appropriately chosen auxiliary variable, admits a finite partial-fraction representation with simple poles on the negative real axis. Its reciprocal admits an analogous representation with a possible additional pole at the origin. Inverting the corresponding Laplace transforms gives finite fractional Prony representations of the relaxation modulus and creep compliance with non-negative amplitudes. These results provide a rigorous proof of a corrected version of the Mainardi--Spada conjecture for this network class. A six-element network illustrates the construction.
\end{abstract}

\newpage
\section{Introduction and Main Results}
\label{sec:intro}
The linear theory of viscoelasticity has been a crucial playground for the study of applications of Fractional Calculus. Indeed, as early as the 1920s, empirical observations by P. G. Nutting \cite{Nutting:1921,Nutting:1943,Nutting:1946} had shown the existence of materials displaying power-law creep. Such materials would display an intermediate behaviour between Hookean elasticity and Newtonian viscous flow. This was then formalized by G. W. Scott--Blair who pioneered the idea of reformulating said empirical power laws in terms of stress-strain relations involving time derivatives of ``fractional'' order \cite{ScottBlair:1944,ScottBlair:1947,ScottBlair:1949}, that was later put in a rigorous framework by M. Caputo and F. Mainardi \cite{Caputo:1966,Caputo:1967,CaputoMainardi:1971a,CaputoMainardi:1971b}. For a detailed historical overview on fractional viscoelasticity we refer the interested reader to \cite{Mainardi:2012} and references therein.

In this work we address a conjecture arising from the formulation of fractional viscoelastic models by F.~Mainardi and G.~Spada in 2011, which can be summarized as follows:
\begin{conjecture}[Mainardi--Spada \cite{Mainardi:2011,Mainardi:2019}]
For $0<\alpha<1$, the fractional viscoelastic models constructed
from elastic Hooke elements (springs) and fractional Scott--Blair
elements (spring-pots) of common order $\alpha$, with positive element
coefficients, and constitutive relations of the form
$$
\left(1+\sum_{k=1}^{p}a_k {\rm D}^{k+\alpha-1}\right)\sigma
=\left(m+\sum_{k=1}^{q}b_k {\rm D}^{k+\alpha-1}\right)\varepsilon \, ,
\qquad m \geq 0 \, , \,\, a_k, b_k \geq 0 \, , \,\, \forall k \, , 
$$
have creep compliance and relaxation modulus
\begin{equation}
    \begin{split}
    J(t)&=J_g+\sum_n J_n
    \left[1 - E_\alpha\left(-(t/\tau_{\varepsilon,n})^\alpha\right)\right]
    +J_+ \, \frac{t^\alpha}{\Gamma(1+\alpha)} \, , \qquad t>0,\\
    G(t)&=G_e+\sum_nG_n
    E_\alpha \left(-(t/\tau_{\sigma,n})^\alpha\right)
    +G_- \, \frac{t^{-\alpha}}{\Gamma(1-\alpha)} \, ,
    \qquad t>0,
    \end{split}
\end{equation}
with finite sums, non-negative amplitudes
$J_g,J_n,J_+,G_e,G_n,G_-$, and positive characteristic times.
\end{conjecture}
This statement, despite its validation for the standard models of fractional linear viscoelasticity \cite{Mainardi:2011,Mainardi2010book}, has, to our knowledge, not been proved for series-parallel networks composed of an arbitrary, yet finite, number of springs and spring-pots of a common order. The aim of this work is therefore to establish a corrected version of this conjecture by proving the following two statements:
\begin{theorem} \label{th-main}
    Let $0<\alpha<1$.
    Consider a fractional viscoelastic model corresponding to a finite series-parallel network of positive springs and spring-pots of order $\alpha$. Let $R(z)$ denote complex shear modulus of the model with respect to the auxiliary variable $z = s^\alpha$. Then the material functions admit the finite {\em fractional Prony representations} 
    \begin{equation}    
    G(t)=G_e+\sum_{j=1}^{N}G_j \, E_\alpha(-\lambda_jt^\alpha)+\frac{G_-t^{-\alpha}}{\Gamma(1-\alpha)} \, ,
    \end{equation}
    \begin{equation}
    J(t)=J_g+\sum_{k=1}^{M}J_k \, \bigl[1-E_\alpha(-\rho_kt^\alpha)\bigr]
    +\frac{J_+t^\alpha}{\Gamma(1+\alpha)} \, ,
    \end{equation}
    for some $M,N \in \N \cup \{0\}$,  $G_e,G_-,J_g,J_+,G_j,J_k\ge0$, relaxation times $\tau_{\sigma,j}=\lambda_j^{-1/\alpha}>0$, and retardation times $\tau_{\varepsilon,k}=\rho_k^{-1/\alpha}>0$; where 
    $$ E_\alpha(z):=\sum_{k=0}^{\infty}\frac{z^k}{\Gamma(\alpha k+1)}, \qquad z\in \mathbb{C} \, ,
    $$
    denotes the one-parameter Mittag-Leffler function.
\end{theorem}

\begin{prop} \label{prop-1}
    Under the assumptions of Theorem~\ref{th-main}, the complex shear modulus $R(z)$ is a non-zero real rational function of the form
    $$
    R(z) = \frac{Q(z)}{P(z)} \, ,
    $$
    where $Q,P \in \R [z]$ are coprime polynomials such that $P(0) = 1$, $Q(0) = m \geq 0$, $p:={\rm deg} (P)$, and $q:= {\rm deg} (Q) \in \{p , p+1\}$. Thus, the constitutive equation associated to the causal network constructed as in Theorem~\ref{th-main} then takes the form
    \begin{equation} \label{eq:const}
    P({\rm D}^{\alpha}) \, \sigma := \left(1+\sum_{k=1}^{p}a_k {\rm D}^{\alpha k}\right)\sigma =\left(m+\sum_{k=1}^{q}b_k {\rm D}^{\alpha k}\right)\varepsilon =: Q ({\rm D}^{\alpha}) \, \varepsilon \, ,
    \end{equation}
    where ${\rm D}^\alpha$ denotes the causal fractional derivative of order $\alpha$, with $a_k>0$ for $1\leq k\leq p$ and $b_k>0$ for $1\leq k\leq q$.
\end{prop}

\begin{remark}
    Eq.~\eqref{eq:const} differs from what was originally conjectured in \cite{Mainardi:2011}. 
    It turns out that the correct order of the operators is $\alpha k$ 
    instead of $k + \alpha -1$. Constitutive equations involving derivative orders different from the conjectured prescription also occur, for instance, in \cite{CostaRibeiro:2012, OkukaZorica:2018}. Furthermore, finite Mittag-Leffler representations and exact common-order interconversion for generalized fractional Maxwell models and generalized fractional Kelvin models are discussed in \cite{KatichaFlintsch:2012}, whereas the present results
    apply to arbitrary finite series--parallel networks.
\end{remark}

The manuscript is organized as follows. In Section~\ref{sec:pre} we provide some generalities on causal distributions and the network composition rules for viscoelastic models. In Section~\ref{sec:lemmas} we collect a series of technical lemmas required for the proofs of the main results. In Sections~\ref{sec:proof-th} and~\ref{sec:proof-prop} we prove the main results of this work. Section~\ref{sec:example} illustrates them with a six-element network. Finally, in Section~\ref{sec:conc} we present some concluding remarks.

\section{Preliminaries}
\label{sec:pre}
We shall work with Laplace-transformable causal distributions supported in $[0,+\infty)$ (see \cite{Widder1946laplace,Doetsch:1974, Zemanian:2003}). 

Let $f,g \in L^1 _{\rm loc} ([0,+\infty))$ such that $f(t)=g(t) = 0$ for all $t < 0$, these functions are therefore well-defined causal distributions. We denote by
$$
(f \ast g) (t) := \int_0 ^t f(t-\tau) g(\tau) \, \d \tau \, ,  
$$
the Laplace (causal) convolution of $f$ and $g$. Furthermore, defining the Laplace transform of $f$ as
$$
\mathcal{L} (f) (s) \equiv \widetilde{f} (s) := \int_0 ^\infty f(t) \, {\rm e}^{-st} \, \d t \, ,
$$
and its extension to causal distributions \cite{Zemanian:2003}, then we can recall the following results
\begin{equation}
\mathcal{L} (f \ast g) (s) = \widetilde{f}(s) \, \widetilde{g}(s) \quad \mbox{and} \quad \mathcal{L} ({\rm D} f) (s) = s \,  \widetilde{f}(s) \, ,
\end{equation}
where ${\rm D}$ denotes the distributional derivative.

Denoting by $H$ the Heaviside unit-step distribution, we introduce the so-called Gelfand--Shilov \cite{MainardiGorenflo1997FC} kernel as
\begin{equation}
\phi _\alpha (t) :=\frac{t^{\alpha-1}}{\Gamma (\alpha)} \, H (t)   \, , \qquad \alpha > 0 \, . 
\end{equation}
Note that, after analytic continuation in $\alpha$ in the sense of distributions, if $\alpha \to -n$, for $n \in \mathbb{N} \cup \{0\}$, then $\phi_\alpha \to {\rm D}^n \delta \equiv \delta^{(n)}$ in the weak sense, with $\delta$ denoting the Dirac delta distribution. Therefore we introduce the notation $\phi_{-n} := \delta^{(n)}$.

For all causal distributions, {\em i.e.} $T \in \mathcal{D}' (\mathbb{R})$ such that ${\rm supp} (T) \subseteq [0,+\infty)$, the convolution
\begin{equation}
{\rm D}^\alpha T := \phi_{-\alpha} \ast T \, ,
\end{equation}
is well-defined for every $\alpha \geq 0$, and denotes the {\em causal fractional derivative}. For such an operator, if $T$ is a Laplace-transformable causal distribution, then it holds that \cite{MainardiGorenflo1997FC} 
\begin{equation} \label{LT-D}
\mathcal{L} ({\rm D}^\alpha T) = s^\alpha \, \widetilde{T} \, .    
\end{equation}

In fractional viscoelasticity, the network construction of a constitutive law relies upon two fundamental elements \cite{Mainardi:2011,Mainardi2010book,Mainardi:2019}. First, the elementary spring, that is meant to represent elastic storage of energy, with constitutive law and symbolic representation
\begin{equation}
    \sigma = m \, \varepsilon \quad \longleftrightarrow \quad \overset{m}{\spring} \, ,
\end{equation}
with $m>0$ denoting the elastic modulus of the elementary spring. Second, the Scott-Blair generalized dashpots, often referred to as spring-pot, with constitutive law and symbolic representation
\begin{equation}
    \sigma = \eta _\alpha \, \D^\alpha \varepsilon \quad \longleftrightarrow \quad \overset{\eta_\alpha}{\springpot} \, ,
\end{equation}
with $\alpha \in (0,1)$ and $\eta_\alpha > 0$ the fractional viscosity. Note that, the spring-pot interpolates between the elementary spring behaviour $\alpha = 0$ and the 
Newtonian viscous dashpot $\alpha = 1$. 

Networks are used to combine these  elementary material responses into a more general constitutive law (see {\em e.g.} \cite{Christensen:1982}). Specifically, when the fundamental elements are connected in {\em parallel}, they undergo the same deformation and their forces add, whereas, when connected in {\em series}, they carry the same force and their deformations add. In the Laplace domain the constitutive laws of the elementary spring and of the spring-pot read
\begin{equation}
    \tilde{\sigma} = m \, \tilde{\varepsilon} \qquad \mbox{and} \qquad  
    \tilde{\sigma} = \eta _\alpha \, s^\alpha \,\tilde{\varepsilon} \, ,
\end{equation}
respectively. 

Given a generic network of springs and spring-pots, the corresponding constitutive law in the Laplace domain reads
\begin{equation} \label{eq:const-gen}
    \tilde{\sigma}(s) = R(s^\alpha) \, \tilde{\varepsilon} (s)
\end{equation}
with $R$ a rational function of $s^\alpha$ denoting the complex shear modulus of the network.\footnote{For comparison, $R(s)$ is denoted by $2 \, \widetilde{\mu}(s)$ in the notation of \cite{Mainardi:2011}.}

Introducing the auxiliary variable $z = s^\alpha$, we have that the complex shear modulus of a spring reads $R_{\rm s} (z) = m$, while for a spring-pot element it is given by $R_{\rm sp} (z) = \eta_\alpha \, z$. The modulus of a composite network is obtained recursively by replacing each pair of connected elements or subnetworks with its equivalent modulus
\begin{equation}
R_{\rm parallel}(z)=R_1(z)+R_2(z),
\qquad
R_{\mathrm{series}}(z)
=\left(\frac{1}{R_1(z)}+\frac{1}{R_2(z)}\right)^{-1} \, ,
\end{equation}
where $R_1(z)$ and $R_2(z)$ denote the moduli of the two elements or subnetworks being connected. Note that 
$R_{\rm s} (z)$ and $R_{\rm sp} (z)$ are polynomials in $\R [z]$, hence each connection preserves the rationality of the collective complex shear modulus.
Furthermore, combining together (in series and/or parallel) finitely many subnetworks will continue to yield a rational modulus. This is a known property of series-parallel networks in linear viscoelasticity (see {\em e.g.} \cite{Bland:2016, Tschoegl:1989} and references therein) 

That said, given a network constructed as above, the relaxation modulus $G$ and creep compliance $J$ are, respectively, the stress response to a unit-step strain $\varepsilon=H$ and the strain response to a unit-step stress $\sigma=H$. Therefore, from Eq.~\eqref{eq:const-gen} we have that the Laplace domain representation of the material functions read
\begin{equation} \label{eq:material}
\widetilde J(s)=\frac{1}{s \, R(s^\alpha)},
\quad \mbox{and} \quad
\widetilde G(s)=\frac{R(s^\alpha)}{s} \, .
\end{equation}

\subsection{Technical Lemmas}
\label{sec:lemmas}
In this section, we collect three lemmas needed to prove the main results. These lemmas restate some known results from classical linear viscoelasticity. They apply in the present fractional setting because, under the substitution ($z=s^\alpha$), the complex modulus of a series–parallel model takes the same form as its classical (integer-order) counterpart. We refer the reader to the earlier works \cite{Gross:1953,Marvin:1962, Bland:2016,Tschoegl:1989,HanygaSeredynska:2007,LoyAnderssen:2014,LoyAnderssen:2014SIAM,Loy:2015,Schilling:2012,Hanyga:2018} for the results summarized in the following statements.

We first characterize the complex shear modulus $R(z)$ associated with a series–parallel network. The following Lemma follows from the analytic characterization and closure properties of complete Bernstein functions (see Theorems 2.2, 6.2 and 7.3, and Corollary 7.6 in \cite{Schilling:2012}). Indeed, the elementary moduli $R_{\rm s}(z)=m$ and $R_{\rm sp}(z)=\eta_\alpha z$ are complete Bernstein functions, and this class is preserved under both addition for parallel connections and reciprocal addition followed by inversion for series connections.

\begin{lemma}\label{lem-1}
    Let $\mathcal{H}^+ := \left\{z \in \C  \, : \, {\rm Im}(z) > 0 \right\}$ and 
    $(0,\infty):=\{z\in\mathbb C \, : \, {\rm Im} (z) = 0 \, , \,\, {\rm Re}(z)>0\}$.\\
    Under the assumptions of Theorem~\ref{th-main}, the complex shear modulus $R$ is a non-zero rational function such that: 
    \begin{enumerate} 
        \item $R$ has no poles on $(0,\infty)$, it is also real-valued and
        strictly positive there;
        \item it is holomorphic on $\mathcal{H}^+$;
        \item it has no zero in $\mathcal{H}^+ \cup (0,
        \infty)$;
        \item ${\rm Im} (R(z)) \geq 0$, for $z \in \mathcal{H}^+$; 
        \item $R^{-1}$ is holomorphic on $\mathcal{H}^+$ and ${\rm Im} [1/R(z)] \leq 0$, for $z \in \mathcal{H}^+$.
    \end{enumerate}
\end{lemma}

We next report partial-fraction representations of $R$ and $R^{-1}$, both of which are rational. The following lemma is a consequence of the representation theorems for complete Bernstein and Stieltjes functions (see \cite[Theorems 6.2 and 7.3]{Schilling:2012}). Rationality ensures that each function has only finitely many poles and extends holomorphically across every interval of the negative real axis containing no pole. The Stieltjes inversion formula \cite[Corollary 6.3]{Schilling:2012} therefore implies that the representing measures have finite support, reducing the integral representations to finite sums (see also \cite{LoyAnderssen:2014SIAM}
for related results on interconversion of discrete measures). 

\begin{lemma} \label{lemma-2}
Let $R$ be a non-zero real rational function satisfying
1.--5. in Lemma~\ref{lem-1}. Then there exist
$N \in \mathbb{N} \cup \{0\}$, $a,b \geq 0$, and, for each
$k \in \{1,\ldots,N\}$, constants $\alpha_k,\lambda_k>0$, with pairwise distinct $\lambda_k$, such that
$$
R(z)=a+bz+\sum_{k=1}^{N}\alpha_k\,\frac{z}{z+\lambda_k},
\qquad z\in\mathbb{C}\setminus\{-\lambda_1,\ldots,-\lambda_N\}.
$$
The sum is understood to be zero when $N=0$.
Moreover, there exist $M\in\mathbb{N} \cup \{0\}$,
$c,d\geq0$, and constants $\beta_k,\rho_k>0$ for each
$k\in\{1,\ldots,M\}$, with pairwise distinct $\rho_k$, such that the reciprocal admits
the rational representation
$$
\frac{1}{R(z)}
=c+\frac{d}{z}+\sum_{k=1}^{M}\frac{\beta_k}{z+\rho_k},
\qquad z\in\mathbb{C}\setminus\{0,-\rho_1,\ldots,-\rho_M\},
$$
where the sum is understood to be zero when $M=0$.
If $d=0$, then the term $d/z$ is omitted and
the expansion of the reciprocal holds on
$\mathbb{C}\setminus\{-\rho_1,\ldots,-\rho_M\}$.
\end{lemma}

\begin{remark}
    It is worth to point out that \cite{HanygaSeredynska:2007,Hanyga:2018} extend the analysis to more general scenarios, including anisotropic viscoelasticity, without assuming a finite series--parallel network structure.
\end{remark}

\begin{remark}
    The pole structure and partial-fraction representations
    have classical counterparts in the theory of electrical
    networks (see, {\em e.g.} \cite{VanValkenburg:1960}).
    In classical linear viscoelasticity, the connection between
    simple poles on the negative real axis and discrete
    relaxation spectra was discussed by \cite{Marvin:1962}.
\end{remark}

Lastly, we recall some useful Laplace transforms.

\begin{lemma}[\cite{Mainardi2010book,Gorenflo2014Mittag-Leffler,giusti2020practicalguide}] \label{lemma-3}
    Let $\alpha \in (0,1)$, $\lambda > 0$, and ${\rm Re} (s) > 0$. The following Laplace transforms hold:
    \begin{eqnarray}
        \mathcal{L} (H) (s) &=& s^{-1} \, , \label{eq:LT-H} \\
        \mathcal{L} (\phi_{1-\alpha}) (s) &=& s^{\alpha - 1} \, , \label{eq:LT-kernel1}\\
        \mathcal{L}(\phi_{1+\alpha})(s) &=& s^{-1-\alpha} \, , \label{eq:LT-kernel2}\\
        \mathcal{L}\!\left[H(t) \, E_\alpha (-\lambda t^\alpha)\right](s) &=&\frac{s^{\alpha-1}}{s^\alpha+\lambda} \, ,
    \end{eqnarray}
    where all complex powers are taken on the principal branch, and the factor $H$ in the last expression indicates extension by zero to $t<0$.
\end{lemma}

\section{Proof of Theorem~\ref{th-main}}
\label{sec:proof-th}
To prove this result we employ standard techniques of linear viscoelasticity  described, for instance, in \cite[Ch. 2, Sec. 5 and 6]{Bland:2016}.

As discussed in Section~\ref{sec:pre}, a generic series-parallel network of springs and spring-pots admits a constitutive relation that, in the Laplace domain, reads as in Eq.~\eqref{eq:const-gen}, with $R \in \R (s^\alpha)$ the complex shear modulus of the network.

By Lemma~\ref{lem-1}, setting $z = s^\alpha$, $R(z)$ satisfies the properties 1.--5. and, by Lemma~\ref{lemma-2}, it admits the representation
$$
R(s^\alpha)=a+bs^\alpha+\sum_{k=1}^{N}\alpha_k\,\frac{s^\alpha}{s^\alpha+\lambda_k} \, ,
\qquad s^\alpha \in\mathbb{C}\setminus\{-\lambda_1,\ldots,-\lambda_N\} \, ,
$$
with $N \in \mathbb{N} \cup \{0\}$, $a,b \geq 0$, and, $\alpha_k,\lambda_k>0$, with pairwise distinct $\lambda_k$, $\forall k \in \{1,\ldots,N\}$.

From \eqref{eq:material} we have that
$$
\widetilde{G} (s) = \frac{R(s^\alpha)}{s} = \frac{a}{s}+bs^{\alpha-1}+\sum_{k=1}^{N}\alpha_k\,\frac{s^{\alpha-1}}{s^\alpha+\lambda_k} \, ,
$$
thus for ${\rm Re} (s) > 0$, $\alpha \in (0,1)$, and considering the principal branch all complex powers, we can apply the results in Lemma~\ref{lemma-3}, which allows us to compute the inverse Laplace transform of $\widetilde{G} (s)$, that reads
$$
G(t) = a H(t) + b \, \phi_{1 - \alpha} (t) + \sum_{k=1}^{N}\alpha_k \, 
H(t) \, E_\alpha (-\lambda_k t^\alpha) \, .
$$
Hence, making the identifications
$$
a \equiv G_e \geq 0 \, , \quad b \equiv G_{-} \geq 0 \, , \quad \alpha_k \equiv G_k > 0 \, , \quad \lambda_k \equiv \tau_{\sigma,k} ^{-\alpha} > 0 \, ,
$$
yields
$$
G(t) = G_e + G_- \, \frac{t^{-\alpha}}{\Gamma(1-\alpha)} 
+ \sum_{k=1}^{N} G_k \, E_\alpha [-(t/\tau_{\sigma,k})^\alpha] \, , \qquad t > 0 \, .
$$

Similarly, from \eqref{eq:material} and by Lemma~\ref{lemma-2} we have that 
$$
\widetilde{J}(s) = \frac{1}{sR(s^\alpha)} =
\frac{c}{s}+\frac{d}{s^{\alpha+1}}+\sum_{k=1}^{M}\frac{\beta_k}{s(s^\alpha+\rho_k)} \, ,
\qquad s^\alpha\in\mathbb{C}\setminus\{0,-\rho_1,\ldots,-\rho_M\} \, ,
$$
with $M\in\mathbb{N} \cup \{0\}$, $c,d\geq0$, and $\beta_k,\rho_k>0$, with pairwise distinct $\rho_k$, $\forall k \in \{1,\ldots,M\}$. Again, for ${\rm Re} (s) > 0$, $\alpha \in (0,1)$, considering the principal branch all complex powers, observing that 
$$
\frac{\beta_k}{s(s^\alpha+\rho_k)} = \frac{\beta_k}{\rho_k} \left( \frac{1}{s} - \frac{s^{\alpha-1}}{s^\alpha+\rho_k} \right) \, , 
$$
and taking advantage of the results in Lemma~\ref{lemma-3}, yields
$$
J(t) = c \, H(t)+d \, \phi _{1+\alpha} (t)+
\sum_{k=1}^{M}
\frac{\beta_k}{\rho_k} 
\, H(t) \, \left[ 1 - E_\alpha (-\rho_k \, t^\alpha) \right] \, ,
$$
hence, making the identifications
$$
c \equiv J_g \geq 0 \, , \quad d \equiv J_+ \geq 0 \, , \quad \frac{\beta_k}{\rho_k} \equiv J_k > 0 \, , \quad \rho_k \equiv \tau_{\varepsilon,k} ^{-\alpha} > 0 \, ,
$$
we get
$$
J(t) = J_g + J_+ \, \frac{t^\alpha}{\Gamma(1+\alpha)} + \sum_{k=1}^{M} J_k \, \left[1-E_\alpha\left(-(t/\tau_{\varepsilon,k})^\alpha\right)\right] \, , \qquad t > 0 \, .
$$

\section{Proof of Proposition~\ref{prop-1}}
\label{sec:proof-prop}
To prove this result, we follow the standard approach described, for instance, in \cite[Ch. 2, Sec. 8]{Bland:2016}.

By the assumptions of Theorem~\ref{th-main}, in view of Lemma~\ref{lemma-2}, the complex shear modulus $R(z)$ admits the representation of the form
$$
R(z)=a+bz+\sum_{k=1}^{N}\alpha_k\,\frac{z}{z+\lambda_k} \, ,
\quad a,b\geq0,\quad \alpha_k,\lambda_k>0 \, ,
$$
with pairwise distinct $\lambda_k$, $\forall k \in \{1,\ldots,N\}$. Moreover, $R \not \equiv 0$, it is regular at zero, and $R(0) = a \geq 0$. 

Since $R \in \R (z)$ we can write $R (z) = \mathcal{N}(z)/{\mathcal{D}(z)}$ with $\mathcal{N},\mathcal{D} \in \R [z]$ coprime polynomials. Furthermore, the regularity of $R$ at $z = 0$, together with coprimality, implies that $\mathcal{D}(0) \neq 0$.

Defining $P(z) := \mathcal{D}(z)/\mathcal{D}(0)$ and $Q(z) := \mathcal{N}(z)/\mathcal{D}(0)$, we have that
$$
R(z) = \frac{Q(z)}{P(z)} \, , \qquad P(0) = 1 \, , \quad Q(0) = R(0) =: m \geq 0 \, .
$$
Let $p:={\rm deg}(P)$ and $q:={\rm deg}(Q)$, the representation gives 
$$
R(x)=bx+ O(1),
\qquad x\to+\infty \, ,
$$
for $z = x \in (0,+\infty)$. If $b>0$, then $R(x)/x \to b >0$, hence comparing this with the leading-order behaviour of $Q(x)/P(x) \sim C \, x^{q-p}$, with $C \neq 0$, yields $q-p=1$. Otherwise, if $b = 0$ then $
R(x) \to a+\sum_{j=1}^{N}\alpha_j>0$ in said limit. This limit could vanish only if $a=0$ and $N=0$, which, together with $b=0$, would imply $R\equiv0$. Thus, if $b = 0$ then $q-p = 0$.

We have therefore proven that $q \in \{ p, p+1\}$, and hence
$$
P(z)=1+\sum_{j=1}^{p}a_j z^j,
\qquad
Q(z)=m+\sum_{k=1}^{q} b_k z^k,
$$
with real coefficients. Furthermore, $a_j,b_k>0$, for all $j\in\{1,\ldots,p\}$
and $k\in\{1,\ldots,q\}$. Indeed, from the representation
of the complex shear modulus $R(z)$ we have that
$$
R(z)=a+bz+\sum_{k=1}^{N}\alpha_k\,\frac{z}{z+\lambda_k}
=\frac{Q(z)}{P(z)}
\quad\Rightarrow\quad
P(z)=\prod_{k=1}^{N}\left(1+\frac{z}{\lambda_k}\right),
$$
where the $\lambda_k$ are distinct, with $P(z)$ being
identified with the normalized reduced denominator
resulting from the addition of the $N$ fractions by the
common denominator procedure. Indeed, no cancellation
occurs, since each $-\lambda_k$ is a pole with non-zero
residue $-\alpha_k\lambda_k$, and hence $p=N$.
Note that, since the coefficients of each linear factor
$\left(1+\frac{z}{\lambda_k}\right)$ are positive,
the polynomial resulting from their product will have
strictly positive coefficients at every degree from $0$
to $p$. Similarly, we have that
\begin{equation*}
\begin{split}
Q(z) &= R(z) P(z) = a P(z)+bz P(z)+\sum_{k=1}^{N}\alpha_k \, z\,\frac{P(z)}{z+\lambda_k}  \\
&= aP(z)+bzP(z)
+\sum_{k=1}^{N}\frac{\alpha_k}{\lambda_k}\,
z\prod_{\ell\neq k}\left(1+\frac{z}{\lambda_\ell}\right) \, .
\end{split}
\end{equation*}
All terms have non-negative coefficients. For $N\geq1$,
the last sum has strictly positive coefficients at every
degree from $1$ to $N$, while, if $b>0$, the term $bzP(z)$ yields
a polynomial of degree $N+1$ with strictly positive coefficients. Therefore $b_k>0$ for all $k\in\{1,\ldots,q\}$. If $N=0$, then
$Q(z)=a+bz$, which is a polynomial of degree at most one with non-negative real coefficients.

Lastly, consider causal, Laplace-transformable stress and strain. In a common half-plane of convergence contained in ${\rm Re} (s)>0$, the constitutive relation of the network reads
$$
\widetilde{\sigma}(s)
=R(s^\alpha) \, \widetilde{\varepsilon}(s) \, .
$$
For ${\rm Re}(s)>0$, write $s=|s|e^{i\theta}$,
where $\theta:={\rm Arg}(s)\in(-\pi,\pi)$ is the principal
argument. Since ${\rm Re}(s)=|s|\cos\theta>0$,
we have $\cos\theta>0$, which within $(-\pi,\pi)$
implies $\theta\in(-\pi/2,\pi/2)$.
Hence, on the principal branch, for $\alpha\in(0,1)$,
$$
{\rm Re}(s^\alpha)
=|s|^\alpha\,\cos(\alpha\theta)>0,
\qquad
|\alpha\theta|<\frac{\alpha\pi}{2}<\frac{\pi}{2} \, .
$$
Since ${\rm Re}(s^\alpha)>0$ for ${\rm Re}(s)>0$, and all poles of $R$ are strictly negative (Lemma~\ref{lemma-2}), the reduced denominator is such that $P(s^\alpha)\neq0$ in that region. Therefore, we can rewrite our Laplace-domain constitutive law as
$$
P(s^\alpha) \, \widetilde{\sigma}(s)
=Q(s^\alpha) \, \widetilde{\varepsilon}(s).
$$
Inverting back to the time domain taking advantage of \eqref{LT-D} and the linearity of the Laplace transform, we find
$$
P({\rm D}^{\alpha})\,\sigma
:=\left(1+\sum_{k=1}^{p} a_k \, 
{\rm D}^{\alpha k}\right)\sigma
=\left(m+\sum_{k=1}^{q} b_k \, 
{\rm D}^{\alpha k}\right)\varepsilon
=:Q({\rm D}^{\alpha})\,\varepsilon \, .
$$
The operator notation is justified by the semigroup property
$({\rm D}^{\alpha})^k={\rm D}^{\alpha k}$ (see \cite{MainardiGorenflo1997FC,Mainardi2010book}).

\section{Example: a six-element network}
\label{sec:example}
\begin{figure}
    \centering
    \includegraphics[width=0.7\linewidth]{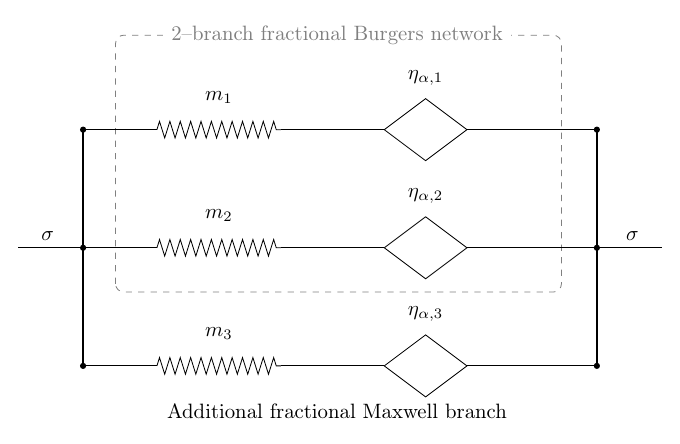}
    \caption{Schematic representation of a six-element network consisting of three fractional Maxwell models of common order $\alpha \in (0,1)$ connected in parallel.}
    \label{fig:1}
\end{figure}

Consider the fractional Burgers network obtained by joining in parallel two fractional Maxwell elements (see {\em e.g.} \cite{Mainardi:2011}). We then enlarge this network by connecting it in parallel to a third, independent fractional Maxwell line. Schematically, this scenario is represented by the network diagram in Figure~\ref{fig:1}. Note that this example is a sub-case of the framework in \cite{CostaRibeiro:2012}, under the assumption of common-order for each fractional Maxwell element.

From standard fractional viscoelasticity
\cite{Mainardi:2011,Mainardi:2019,Mainardi2010book},
we know that each fractional Maxwell branch in
Figure~\ref{fig:1} is governed by a constitutive equation
that reads
\begin{equation}
    \sigma_j+\frac{\eta_{\alpha,j}}{m_j}\,
    {\rm D}^{\alpha}\sigma_j
    =\eta_{\alpha,j}\,{\rm D}^{\alpha}\varepsilon \, ,
    \qquad m_j>0 \, , \quad \eta_{\alpha,j}>0 \, ,
    \quad \mbox{for } j=1,2,3 \, .
\end{equation}
When such branches are connected in parallel, they share
the same strain history $\varepsilon$, while their stress
histories add up
(see \cite{Christensen:1982,Mainardi2010book}).
Hence the total stress history of the network reads
\begin{equation}
    \sigma=\sigma_1+\sigma_2+\sigma_3 \, .
\end{equation}
Computing the total stress history of the network is much easier in the Laplace domain, indeed there the constitutive law of each branch reads
\begin{equation}
    \left(1 + \frac{\eta_{\alpha,j}}{m_j}\,
    s^{\alpha} \right) \widetilde{\sigma}_j
    =\eta_{\alpha,j}\,s^\alpha \, \widetilde{\varepsilon} \, ,
    \qquad m_j>0 \, , \quad \eta_{\alpha,j}>0 \, ,
    \quad \mbox{for } j=1,2,3 \, ,
\end{equation}
and hence the total stress history is given by
\begin{equation}
    \widetilde{\sigma} = \sum_{j=1}^3 \widetilde{\sigma}_j = 
    \left( \sum_{j=1}^3 \eta_{\alpha,j} \, \frac{s^\alpha}{1 + \frac{\eta_{\alpha,j}}{m_j}\,
    s^{\alpha}} \right) \, \widetilde{\varepsilon} \, ,
\end{equation}
which yields a complex shear modulus that reads
\begin{equation} \label{eq:R6}
    R(z) = \sum_{j=1}^3 m_j \, \frac{z}{z+\lambda_j},
\qquad \lambda_j=\frac{m_j}{\eta_{\alpha,j}}>0 \, , \qquad z \equiv s^\alpha \, . 
\end{equation}
Hence, using Lemma~\ref{lemma-3}, the corresponding relaxation modulus reads
\begin{equation}
    G(t) = \mathcal{L}^{-1} \left[\frac{R(s^\alpha)}{s} \right] (t)= \sum_{j=1}^3 m_j \, E_\alpha \left(- t^\alpha / \tau^\alpha _{\sigma,j} \right) \, , \qquad \tau^\alpha _{\sigma,j} := \frac{1}{\lambda_j} = \frac{\eta_{\alpha,j}}{m_j} \, , \qquad t > 0 \, .
\end{equation}

In what follows, we assume that $0<\lambda_1<\lambda_2<\lambda_3$. If some of the $\lambda_j$ coincide, the corresponding terms in \eqref{eq:R6} can be combined by adding their moduli, thereby reducing the number of terms in the sum.

\begin{figure}
    \centering
    \includegraphics[width=0.75\linewidth]{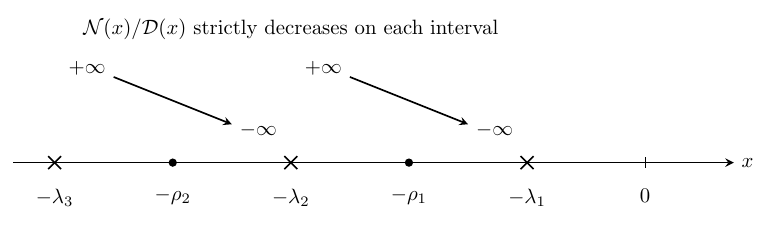}
    \caption{Poles (crosses) and zeros (dots) of $\mathcal{N} (x)/\mathcal{D} (x)$ on the real axis. The zero positions are schematic and assume the ordering $-\lambda_3<-\lambda_2<-\lambda_1$. The arrows above the axis schematically show the decrease of $\mathcal{N} (x)/\mathcal{D} (x)$ from $+\infty$ to $-\infty$.}
    \label{fig:2}
\end{figure}

To compute the creep compliance of the model we have to study the reciprocal of the complex shear modulus. First, performing the sum in $R(z)$ we get
\begin{equation}
    R(z) = z \, \frac{m_1 (z + \lambda_2)(z + \lambda_3) + m_2 (z + \lambda_1)(z + \lambda_3) + m_3 (z + \lambda_1)(z + \lambda_2)}{(z + \lambda_1)(z + \lambda_2)(z + \lambda_3)} \equiv z \, \frac{\mathcal{N} (z)}{\mathcal{D} (z)} \, .
\end{equation}
The numerator of $R(z)/z$ can be written as
\begin{equation}
    \mathcal{N} (z) = a z^2 + b z + c \, ,
\end{equation}
with
\begin{equation}
    \begin{split}
    a&=m_1+m_2+m_3 \, ,\\
    b&=m_1(\lambda_2+\lambda_3)
  +m_2(\lambda_1+\lambda_3)
  +m_3(\lambda_1+\lambda_2) \, ,\\
    c&=m_1\lambda_2\lambda_3
  +m_2\lambda_1\lambda_3
  +m_3\lambda_1\lambda_2 \, .
    \end{split}
\end{equation}
Furthermore, observing that \footnote{This expansion of $R(z)/z$ is akin to that of Eq.~(70) in \cite{TudelaPiColombaro:2026}, yet the decomposition presented here is exact.}
\begin{equation}
    \frac{\mathcal{N} (z)}{\mathcal{D} (z)} = \frac{R(z)}{z} = \sum_{j=1}^{3}\frac{m_j}{z+\lambda_j} \, ,
\end{equation}
then taking $z = x \in \R$ we have that 
$$
\frac{\d }{\d x} \left( \frac{\mathcal{N} (x)}{\mathcal{D} (x)}\right) =
-\sum_{j=1}^{3}\frac{m_j}{(x+\lambda_j)^2}<0 \, ,
$$
on each real interval between consecutive poles. Hence, the function is strictly monotonically decreasing from  $+\infty$ to $-\infty$ on each of these intervals, therefore $\mathcal{N} (x)/\mathcal{D} (x)$ has exactly one zero on each interval. We can therefore conclude that $\mathcal{N} (x)/\mathcal{D} (x)$ has two distinct zeros at $-\rho_1$ and $-\rho_2$ on the negative real line, and we conventionally take $-\rho_2 < -\rho_1$. This is schematically represented in Figure~\ref{fig:2}. We can therefore factorize $\mathcal{N} (z)$ as 
\begin{equation}
    \mathcal{N} (z) = a (z+ \rho_1)(z+ \rho_2) \, , \qquad \rho_{1,2} = -\frac{-b \pm \sqrt{b^2 - 4ac}}{2a} \, ,
\end{equation}
and $\rho_{1,2}$ are strictly positive as discussed above. 

The reciprocal of the complex shear modulus then reads
\begin{equation} \label{eq:1overR6}
    \frac{1}{R(z)} = \frac{1}{z} \frac{\mathcal{D} (z)}{\mathcal{N}  (z)} = \frac{1}{z} \frac{(z + \lambda_1)(z + \lambda_2)(z + \lambda_3)}{a (z+ \rho_1)(z+ \rho_2)} \, .
\end{equation}
Since both zeros and poles of $[R(z)]^{-1}$ are simple and both the numerator and the denominator are polynomials of the same order, we have that
\begin{equation}
\frac{1}{R(z)}
= A +\frac{B}{z}+\frac{C}{z+\rho_1}
+\frac{D}{z+\rho_2} \, ,
\end{equation}
for some $A,B,C,D \in \R$.

The constant $A$ is obtained as an asymptotic limit on the real line, {\em i.e.}
\begin{equation}
    A = \lim_{\R \ni z \to + \infty}
    \frac{1}{R(z)} = \lim_{\R \ni z \to + \infty}
    \frac{1}{a} \frac{(z + \lambda_1)(z + \lambda_2)(z + \lambda_3)}{z (z+ \rho_1)(z+ \rho_2)} = \frac{1}{a} = \frac{1}{m_1 + m_2 + m_3} \, .
\end{equation}
The coefficients $B,C,D$ instead are simply the residues at $z_0 = 0, -\rho_1, -\rho_2$ of $[R(z)]^{-1}$. Since these poles are all simple, then
\begin{equation}
    {\rm Res} \left( [R(z)]^{-1}, z_0 \right) = \lim_{z \to z_0} (z-z_0) [R(z)]^{-1} \, ,
\end{equation}
which implies
\begin{equation}
    B = \lim_{z\to0}\frac{z}{R(z)}
=\left(\sum_{j=1}^{3}\frac{m_j}{\lambda_j}\right)^{-1}
=\frac1{\eta_{\alpha,1}+\eta_{\alpha,2}+\eta_{\alpha,3}} \, ,
\end{equation}
\begin{equation}
    C = \lim_{z\to-\rho_1}\frac{z + \rho_1}{R(z)} = \frac{1}{R'(-\rho_1)} = 
    \left[
    \sum_{j=1}^{3}
    \frac{m_j\lambda_j}{(\lambda_j-\rho_1)^2}
    \right]^{-1}>0 \, ,
\end{equation}
\begin{equation}
    D = \lim_{z\to-\rho_2}\frac{z + \rho_2}{R(z)} = \frac{1}{R'(-\rho_2)} = 
    \left[
    \sum_{j=1}^{3}
    \frac{m_j\lambda_j}{(\lambda_j-\rho_2)^2}
    \right]^{-1}>0 \, ,
\end{equation}
where we have taken advantage of \eqref{eq:R6} and the fact that
$$
R'(z) = \sum_{j=1}^{3} m_j \, \frac{\d}{\d z}
\left(\frac{z}{z+\lambda_j} \right) =
\sum_{j=1}^{3}\frac{m_j\lambda_j}{(z+\lambda_j)^2} \, .
$$
The creep compliance of the network then reads
\begin{equation}
\widetilde J(s)=\frac1{s \, R(s^\alpha)}
=\frac{A}{s}+\frac{B}{s^{1+\alpha}}
+ \frac{C}{s(s^\alpha+\rho_1)} 
+ \frac{D}{s(s^\alpha+\rho_2)} \, ,
\end{equation}
and hence, using the decomposition
$$
\frac{1}{s(s^\alpha+\rho_k)}
=\frac{1}{\rho_k}
\left(\frac{1}{s}-\frac{s^{\alpha-1}}{s^\alpha+\rho_k}\right) \, ,
$$
inverting back to the time domain (Lemma~\ref{lemma-3}) yields
\begin{equation}
J(t)= A +\frac{B \, t^\alpha}{\Gamma(1+\alpha)}+ 
\frac{C}{\rho_1}\left[1-E_\alpha(-\rho_1 t^\alpha)\right] + 
\frac{D}{\rho_2}\left[1-E_\alpha(-\rho_2 t^\alpha)\right] \, , \qquad t> 0 \, ,
\end{equation}
with $\rho_k = \tau^{-\alpha}_{\varepsilon,k}$ for $k =1,2$.

Lastly, inspecting \eqref{eq:1overR6} it is easy to see that 
$$
P(z) = \frac{(z + \lambda_1)(z + \lambda_2)(z + \lambda_3)}{\lambda_1 \lambda_2 \lambda_3}
\quad \mbox{and} \quad
Q(z) = \frac{a \, z \, (z+ \rho_1)(z+ \rho_2)}{\lambda_1 \lambda_2 \lambda_3} \, ,
$$
and hence the constitutive relation of this six-element network (schematically depicted in Figure~\ref{fig:1}) is of the form
$$
\left(1+\sum_{k=1}^{3}a_k {\rm D}^{\alpha k}\right)\sigma =\sum_{k=1}^{3}b_k {\rm D}^{\alpha k} \varepsilon \, ,
$$
note that $m = R(0) = 0$ in this case.

\begin{figure}
    \centering
    \includegraphics[width=\linewidth]{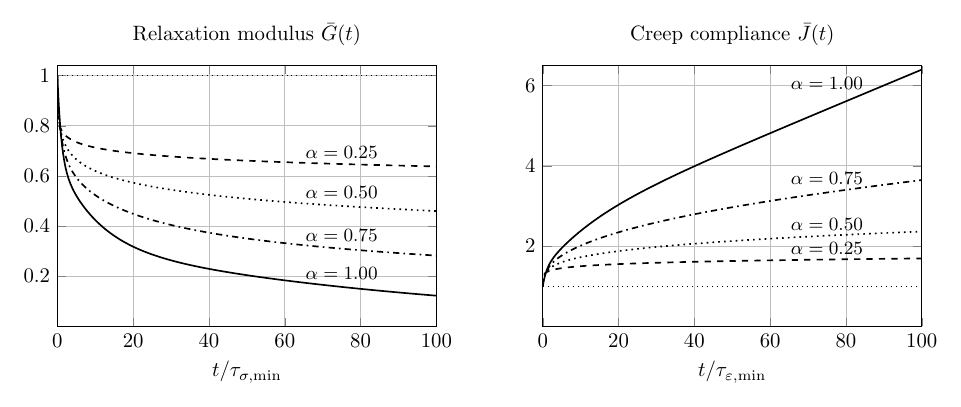}
    \caption{Dimensionless relaxation modulus (left) and creep compliance (right) of the 3-branch fractional Maxwell network.
    The responses are normalized as $\bar{G}(t)=G(t)/M_0$ and $\bar{J}(t)= M_0 \, J (t)$, where $M_0=m_1 + m_2 + m_3$ (thus $\bar G(0^+)=\bar J(0^+)=1$).
    With respect to a common reference time $t_{\ast}$,
    the chosen dimensionless parameters are
    $\bar m_j=m_j/M_0=1/3$ and $(\bar{\lambda}_1,\bar{\lambda}_2,\bar{\lambda}_3)=(0.1,1,10)$, where $\bar{\lambda}_j=t_{\ast}^{\alpha}\lambda_j$. In this case we set $t_\ast=\lambda_2^{-1/\alpha}=\tau_{\sigma,2}$. Consequently, $(\bar{\eta}_{\alpha,1},\bar{\eta}_{\alpha,2},\bar{\eta}_{\alpha,3})=(10/3,1/3,1/30)$, with $\bar{\eta}_{\alpha,j}=\eta_{\alpha,j}/(M_0 \,t_{\ast}^{\alpha})$. These parameters are held fixed for all values of $\alpha$, while the normalizing times $\tau_{\sigma,\min}=(\max_j\lambda_j)^{-1/\alpha}$ and $\tau_{\varepsilon,\min}=(\max_k\rho_k)^{-1/\alpha}$ depend on $\alpha$.}
    \label{fig:3}
\end{figure}

These results match the general rigorous predictions of Theorem~\ref{th-main} and Proposition~\ref{prop-1}. Some plots of the resulting material functions are reported in Figure~\ref{fig:3}.

\section{Conclusions}
\label{sec:conc}
In this work we have proven that the relaxation modulus and creep compliance of any finite series–parallel network of springs and spring-pots of a common order $0<\alpha<1$ admit finite fractional Prony representations with non-negative amplitudes. To this end, we have applied classical representation results for complete Bernstein and Stieltjes functions to the complex shear modulus expressed in the auxiliary variable $z=s^\alpha$, obtaining finite partial-fraction representations of both the modulus and its reciprocal. These results provide a rigorous proof of a corrected version of the Mainardi--Spada conjecture \cite{Mainardi:2011} for this class of networks. Additionally, we have shown that the corresponding constitutive equation involves derivatives of order $k\alpha$, rather than $k+\alpha-1$ as originally conjectured, with positive non-constant coefficients and polynomial degrees $q-p \in \{0,1 \}$. Lastly, we have illustrated these results through a six-element network consisting of three fractional Maxwell elements connected in parallel.
\section*{Acknowledgements}
\noindent A.G. is supported by the Italian Ministry of Universities and Research (MUR) through the grant ``BACHQ: Black Holes and The Quantum'' (grant no. J33C24003220006) and by the INFN project FLAG. This work has been carried out in the framework of activities of the National Group of Mathematical Physics (GNFM, INdAM). 

\medskip

\noindent {\bf AI Disclosure.} OpenAI's GPT--6 Astra was used during October 2026 for critical feedback, bibliographic assistance, discussion of selected proof steps, and assistance with figure preparation using TikZ. The proof strategy was conceived by the authors. All AI-assisted content was independently checked for correctness by the authors, who retain full responsibility for the manuscript.

\appendix

\section{Pedagogical Appendix}
In this appendix, we derive Lemmas~\ref{lem-1} and~\ref{lemma-2} using elementary complex analysis and the Mittag--Leffler theorem. Versions of these results are well established in the literature on viscoelasticity and related function theory (see \cite{Gross:1953,Marvin:1962, Bland:2016,Tschoegl:1989,HanygaSeredynska:2007,LoyAnderssen:2014,LoyAnderssen:2014SIAM,Loy:2015,Schilling:2012,Hanyga:2018}). Our purpose is pedagogical, and we make no claim of priority for the results themselves.

\medskip

\noindent {\em Proof of Lemma~\ref{lem-1}.} 
It suffices to use the induction principle on the construction of the network. 
    
    First, we observe that for the elementary constituents of the network we have
    $$
    R_{\rm s}(z)=m \quad \mbox{and} \quad R_{\rm sp}(z)=\eta_\alpha z \, ,
    $$
    thus all five properties hold because of the assumption $m, \eta_\alpha > 0$.

    Assume that the five properties hold for two subnetworks with shear moduli $R_1$ and $R_2$, respectively.
\begin{enumerate}
    \item Denoting by $\R (z)$ the field of rational functions in one variable over the real numbers, we have that $R \in \R (z)$. If we then restrict $R|_{(0,\infty)}$ we have that $R|_{(0,\infty)} \in \R (x)$ with $x > 0$, and hence $R|_{(0,\infty)}$ is a real function for both the parallel and series constructions. Furthermore, by the induction hypotheses we have that $R_1(x)$ and $R_2(x)$ are both finite and strictly positive for $x > 0$. Therefore we have that
    $$ 
    R_{\rm parallel}(x)=R_1(x)+R_2(x)>0 \quad \mbox{and} \quad 
    R_{{\rm series}}(x) =\left[(R_1(x))^{-1}+(R_2(x))^{-1}\right]^{-1} >0 \, , 
    $$
    the resulting modulus has no poles on $(0,\infty)$ for either connection.
    \item For a parallel connection the resulting shear modulus reads $R = R_1 + R_2$, and since $R_1,R_2$ are holomorphic on $\mathcal{H}^+$ by the induction assumption, then so is the sum of the two on $\mathcal{H}^+$. If instead we connect the two subnetworks in series the resulting shear modulus is given by $R^{-1} = R_1^{-1} + R_2^{-1}$. Hence, by the induction hypotheses, $R^{-1}$ is also holomorphic on $\mathcal{H}^+$, with ${\rm Im}(R^{-1}(z)) \leq 0$ for $z \in \mathcal{H}^+$. To conclude that $R$ is holomorphic on $\mathcal{H}^+$, we must show that $R^{-1}$ has no zeros there. Suppose that $R^{-1}(z_0)=0$ for some $z_0 \in \mathcal{H}^+$. Since $R^{-1}$ is holomorphic, then ${\rm Im}(R^{-1})$ is harmonic. By the strong maximum principle, the harmonic function ${\rm Im}(R^{-1})$ must vanish identically on $\mathcal{H}^+$. The Cauchy-Riemann equations then imply that $R^{-1}$ is constant, and since $R^{-1}(z_0)=0$, we obtain $R^{-1}\equiv0$ on $\mathcal{H}^+$. However, for $x \in (0,+\infty)$, the induction hypotheses give $R^{-1}(x)=R_1^{-1}(x)+R_2^{-1}(x)>0$. Since $R^{-1}$ is rational and has no pole at $x$, it is continuous there, and therefore
    $$
    R^{-1}(x)=\lim_{\varepsilon\to 0^+} R^{-1}(x+i\varepsilon)=0 \, ,
    $$
    given that $x+i\varepsilon\in\mathcal{H}^+$, where $R\equiv0$. Hence, we have reached a contradiction. Thus $R^{-1}$ has no zeros on $\mathcal{H}^+$, and consequently $R$ is holomorphic there.
    \item For a parallel connection,
    $$
    {\rm Im}(R(z)) = {\rm Im}(R_1(z)) + {\rm Im}(R_2(z)) \geq 0 \, ,
    \qquad \forall z \in \mathcal{H}^+ \cup (0,\infty) \, ,
    $$
    since, by the induction hypotheses, ${\rm Im}(R_1(z)), \, {\rm Im}(R_2(z)) \geq 0$ for $z \in \mathcal{H}^+ \cup (0,\infty)$. Since $R$ is holomorphic on $\mathcal{H}^+$, ${\rm Im}(R)$ is harmonic there. If we assume that $\exists z_0 \in \mathcal{H}^+$ such that $R(z_0)=0$, then ${\rm Im}(R(z_0)) = 0 = \inf_{\mathcal{H}^+}({\rm Im}(R))$. Hence, by the strong minimum principle, ${\rm Im}(R)\equiv0$ on $\mathcal{H}^+$. This means that $R(z) = {\rm Re}(R(z))$, {\em i.e.} $R$ is a real-valued holomorphic function. The Cauchy--Riemann equations then force $R$ to be constant. Since we assumed that $R(z_0)=0$, we obtain $R\equiv0$ on $\mathcal{H}^+$. However, for $x \in (0,+\infty)$, the induction hypotheses give $R(x)=R_1(x)+R_2(x)>0$. Since $R$ is rational and has no pole at $x$, it is continuous there. Therefore,
    $$
    R(x)=\lim_{\varepsilon\to0^+}R(x+i\varepsilon)=0 \, ,
    $$
    since $x+i\varepsilon\in\mathcal{H}^+$, where $R\equiv0$. This contradicts $R(x)>0$, thus proving that $R$ has no zeros in $\mathcal{H}^+$.

    For a series connection, $R=1/(R_1^{-1}+R_2^{-1})$. The denominator is holomorphic and nowhere zero on $\mathcal{H}^+$, as established in the previous step. Consequently, $R$ has no zeros there. 
    
    Finally, for either connection, the induction hypotheses imply $R(x)>0$ for every $x \in (0,+\infty)$, so $R$ has no zeros on the positive real axis either.
    \item The positivity in the case of the parallel connection has already been shown in the previous step. We shall now prove it for the series connection. To this end it suffices to observe that by the induction hypothesis it holds ${\rm Im}(R_1(z)), \, {\rm Im}(R_2(z)) \geq 0$ for $z \in \mathcal{H}^+$. Hence, we have that
    $$
    {\rm Im} (R^{-1}) = {\rm Im} (R^{-1} _1) + {\rm Im} (R^{-1}_2)
    =-\frac{{\rm Im} (R_1)}{|R_1|^2} -\frac{{\rm Im}(R_2)}{|R_2|^2} \leq 0 \, ,
    $$
    on $\mathcal{H}^+$. Then, since by the previous step we know that $[R(z)]^{-1} \neq 0$, $\forall z \in \mathcal{H}^+$ we have that its reciprocal is such that
    $$
    {\rm Im} (R) = {\rm Im} [(R^{-1})^{-1}] = -\frac{{\rm Im} (R^{-1})}{|R^{-1}|^2} \geq 0 \, .
    $$
    \item By 2. and 3., $R$ is holomorphic and nowhere zero on $\mathcal{H}^+$, so $1/R$ is holomorphic there. By 3. we have that
    $$
    {\rm Im} \left[(R(z))^{-1} \right] =-\frac{{\rm Im} (R(z))}{|R(z)|^2} \leq 0 \, ,
    $$
    on $\mathcal{H}^+$. 
\end{enumerate}
\hfill $\square$

\medskip

\noindent {\em Proof of Lemma~\ref{lemma-2}.} 
Upon excluding all possible removable singularities,
and writing $R$ in reduced form, we proceed as follows.

First, by Lemma~\ref{lem-1} we have that $R(z)$ is
holomorphic on $\mathcal{H}^+$, and hence it has no poles
on $\mathcal{H}^+$. Furthermore, since $R \in \R(z)$,
its poles correspond to the zeros of the denominator,
which in turn is a polynomial with real coefficients.
Since a polynomial with real coefficients satisfies
$P(\overline{z})=\overline{P(z)}$, all its non-real zeros
occur in conjugate pairs.
Hence, since $R(z)$ has no poles on $\mathcal{H}^+$,
it will have no poles also in the lower-half plane
$\mathcal{H}^- := \left\{z \in \C \, : \,
{\rm Im}(z)<0\right\}$.
Moreover, since by Lemma~\ref{lem-1} we have that
$R|_{(0,\infty)}$ is both defined and strictly positive,
then $R$ has no poles also on $(0,\infty)$.
Hence, all poles of $R$ lie on
$(-\infty,0] := \{z \in \C \, : \,
{\rm Re}(z)\leq0 \,\, \mbox{and} \,\,
{\rm Im}(z)=0\}$.

Let $x_0\leq0$ be a pole of order $m \in \N$.
Locally, by Laurent's theorem, we have that
$$
R(z)=\frac{a_{-m}}{(z-x_0)^m}
+O\!\left(|z-x_0|^{-m+1}\right),
\qquad a_{-m}\in\R\setminus\{0\}.
$$
Indeed, $(z-x_0)^mR(z)$ has a removable
singularity at $x_0$ and extends holomorphically there,
with non-zero value $a_{-m}$. Its limit is therefore
independent of the direction. Taking this limit along
the real axis, where $R$ is real-valued, gives
$$
a_{-m}
=\lim_{\R\ni x\to x_0}(x-x_0)^mR(x)\in\R \, .
$$

Let us show that the pole is simple, {\em i.e.} $m=1$.
Let $z=x_0+\varepsilon e^{i\theta}$, with
$\theta\in(0,\pi)$ and $\varepsilon>0$,
so that $z\in\mathcal{H}^+$.
Therefore we have that
\begin{equation*}
\begin{split}
R(z)
&=\frac{a_{-m}}{(z-x_0)^m}
+O\!\left(|z-x_0|^{-m+1}\right)\\
&=\frac{a_{-m}}{\varepsilon^m\,\eu^{im\theta}}
+O\!\left(\varepsilon^{1-m}\right).
\end{split}
\end{equation*}
Hence, from $a_{-m}\in\R$ we have that
$$
R(z)
=\frac{a_{-m}}{\varepsilon^m}\,\eu^{-im\theta}
+O\!\left(\varepsilon^{1-m}\right)
\quad\Rightarrow\quad
{\rm Im}[R(z)]
=-\frac{a_{-m}}{\varepsilon^m}\,\sin(m\theta)
+O\!\left(\varepsilon^{1-m}\right).
$$

If $m\geq2$, the function $\sin(m\theta)$ takes both
signs on $(0,\pi)$. In particular, if we fix
$\theta\in(0,\pi)$ such that
$a_{-m}\,\sin(m\theta)>0$, then for that
fixed $\theta$ we have that
${\rm Im}[R(z)]
=-\varepsilon^{-m}\,a_{-m}\,\sin(m\theta)
+O\!\left(\varepsilon^{1-m}\right)<0$
for sufficiently small $\varepsilon$.
This however conflicts with ${\rm Im}[R(z)]\geq0$
on $\mathcal{H}^+$, hence we have reached a contradiction.
Therefore it must be $m=1$, namely $R(z)$ has only
simple poles on $(-\infty,0]$.

It is also important to observe that if
$z=x_0\in(-\infty,0]$ is a pole of $R(z)$,
then its residue is negative.
Indeed, we have proven that such a pole can only
be simple, which means that by Laurent's theorem,
we have that
$$
R(z)=\frac{a_{-1}}{z-x_0}+O\!\left(1\right),
$$
hence the condition ${\rm Im}[R(z)]\geq0$
on $\mathcal{H}^+$ implies, for
$z=x_0+\varepsilon e^{i\theta}$
with $\theta\in(0,\pi)$, that
$$
{\rm Im}[R(z)]
=-\frac{a_{-1}}{\varepsilon}\,\sin(\theta)
+O\!\left(1\right)\geq0.
$$
Since $\sin\theta>0$ on $(0,\pi)$,
multiplying by $\varepsilon$ and letting
$\varepsilon\to0^+$ gives $a_{-1}\leq0$.
Since the pole is simple, $a_{-1}\neq0$,
and therefore $a_{-1}<0$.
But by definition $a_{-1}:={\rm Res}(R,x_0)$,
and hence the residue of $R(z)$ at a pole is
necessarily negative.

These results allow us to show that $z=0$
can never be a pole for $R(z)$.
Indeed, suppose, by way of contradiction,
that $z=0$ is a pole.
Then it must be a simple pole with negative residue
by the arguments above.
This means that by Laurent's theorem we have
$$
R(z)=-\frac{|a_{-1}|}{z}+O\!\left(1\right),
\qquad z\to0 \, ,
$$
and, in particular,
$$
R(x)=-\frac{|a_{-1}|}{x}+O\!\left(1\right),
\qquad x\to0^+ \, .
$$
By Lemma~\ref{lem-1}, $R(x)>0$ for all $x>0$,
however for sufficiently small $x>0$ we have that
$R(x)=-|a_{-1}|/x+O\!\left(1\right)<0$,
and hence we have reached a contradiction.
Therefore $z=0$ can never be a pole of $R(z)$.

Let us now recall a well-known theorem by
G.~Mittag--Leffler
(see {\em e.g.} \cite[Ch.~VIII]{Conway:1978}):
Let $A\subseteq\C$ be open, and let $\{z_k\}$ be a
sequence of distinct points in $A$ without a limit
point in $A$. For each $k$, let
$$
p_k(z)=\sum_{j=1}^{m_k}\frac{c_j^{(k)}}{(z-z_k)^j},
\qquad m_k\geq1,\quad c_{m_k}^{(k)}\neq0 \, .
$$
Then there exists a meromorphic function $f$ on $A$
whose poles are precisely the points $z_k$
and whose singular part at $z_k$ is $p_k(z)$.
Any two such functions differ by a holomorphic
function on $A$.

In our case, $R$ is rational and therefore
meromorphic on $\C$, with finitely many poles
$z_k=x_k<0$, all of which are simple and with
a negative residue.
Thus, in the finite-pole case of Mittag--Leffler's
theorem, we may write
$$
R(z)=h(z)+\sum_{k=1}^{N}p_k(z),
\qquad
p_k(z)=\frac{{\rm Res}(R,x_k)}{z-x_k},
$$
where $h$ is an entire function and $N$ is the
number of (simple) poles of $R$.
Moreover, since
$h(z)=R(z)-\sum_k{\rm Res}(R,x_k)/(z-x_k)$,
it is the difference of two rational functions
with real coefficients, and hence $h\in\R(z)$.
Furthermore, since $h$ is both rational and entire,
it must be a polynomial with real coefficients.
Otherwise, its reduced denominator would be
non-constant and, by the fundamental theorem
of algebra, would have a zero, yielding a pole of $h$.

Let us show that $h\in\R[z]$ is a polynomial
of degree at most one, {\em i.e.}
$h(z)=\hat{a}+b\,z$ with $\hat{a},b\in\R$.
Indeed, consider $z=r\,\eu^{i\theta}$
with $r>0$ and $\theta\in(0,\pi)$ so that $z\in\mathcal{H}^+$
and assume ${\rm deg}(h)=:m\geq2$, then
$$
h(z)=c_m\,z^m+\cdots+c_1\,z+c_0,
\qquad c_j\in\R,\quad
\forall j\in\{0,\ldots,m\},\quad c_m\neq0.
$$
For each fixed $\theta\in(0,\pi)$,
the finite sum of the principal parts satisfies
$$
\sum_{k=1}^{N}p_k(r\,\eu^{i\theta})=O(r^{-1}) \, ,
\qquad r\to+\infty \, ,
$$
and hence
$$
{\rm Im}(R(z))
=c_m\,r^m\,\sin(m\theta)+O(r^{m-1}).
$$
Choosing a fixed $\theta\in(0,\pi)$ such that
$c_m\sin(m\theta)<0$, then for a sufficiently large
$r$ we have that ${\rm Im}(R(z))<0$
within $\mathcal{H}^+$, thus conflicting with
the non-negativity of that quantity on
$\mathcal{H}^+$.
Hence we have reached a contradiction.
Therefore $h(z)=\hat{a}+b\,z$,
including the possibility that $h$ is identically zero.

Next, let us prove that $b\geq0$.
Let $h(z)=\hat{a}+b\,z$ with $\hat{a},b\in\R$
and let $z=ir$ with $r>0$, then
${\rm Im}[R(ir)]=b\,r+O(r^{-1})\geq0$
for large $r$.
Hence, to guarantee ${\rm Im}[R(ir)]\geq0$
as $r\to+\infty$, it must be that $b\geq0$.

Lastly, setting
$\lambda_k:=-x_k=|x_k|>0$
and $\alpha_k:=-{\rm Res}(R,x_k)/|x_k|$
so that $\alpha_k>0$,
and observing that
$$
\frac{{\rm Res}(R,x_k)}{z-x_k}
=\alpha_k\frac{z}{z+|x_k|}-\alpha_k,
$$
we obtain
\begin{equation*}
\begin{split}
R(z)
&=h(z)+\sum_{k=1}^{N}p_k(z)
=\hat{a}+bz+\sum_{k=1}^{N}
\frac{{\rm Res}(R,x_k)}{z-x_k}\\
&=\hat{a}-\sum_{k=1}^{N}\alpha_k
+bz+\sum_{k=1}^{N}\alpha_k\frac{z}{z+|x_k|}.
\end{split}
\end{equation*}
Denoting $a:=\hat{a}-\sum_{k=1}^{N}\alpha_k=R(0)$,
since $R$ is regular at zero and positive on
$(0,\infty)$, then
$$
a=\lim_{\R\ni x\to0^+}R(x)\geq0.
$$
This concludes the proof of the first part
of the statement.

By the assumptions, the reciprocal $R^{-1}$
is real rational, holomorphic on $\mathcal{H}^+$,
positive on $(0,\infty)$, and satisfies
${\rm Im}(R^{-1})\leq0$.
The same local argument with the sign reversed
shows that its poles are simple and non-positive,
with positive residues.
Mittag--Leffler's theorem and rationality give
$(R(z))^{-1}
=\ell(z)+d/z+\sum_{k=1}^{M}\beta_k/(z+\rho_k)$,
where $\ell$ is a real polynomial, $d\geq0$,
and $\beta_k,\rho_k>0$.
The sign of ${\rm Im}(R^{-1})$ forces
$\ell(z)=c+e\,z$ with $e\leq0$,
while positivity on $(0,\infty)$ excludes $e<0$.
Indeed, if $e<0$, then
$$
\frac{1}{R(x)}=ex+c+O(x^{-1})<0
\qquad \mbox{for sufficiently large } x>0 \, .
$$
Thus $e=0$ and
$$
c=\lim_{x\to+\infty}\frac{1}{R(x)}\geq0.
$$
This yields the required expansion, with the term
$d/z$ omitted when $d=0$. \hfill $\square$

\bibliographystyle{utphys}
\bibliography{ref}
\end{document}